\documentclass[twocolumn,preprintnumbers,amsmath,amssymbm,prl]{revtex4}
\usepackage{epsfig}
\usepackage{graphicx}
\usepackage{amssymb}

\begin{document}
\title{Cosmic Censorship: Formation of a Shielding Horizon Around a Fragile Horizon}
\author{Shahar Hod}
\address{The Ruppin Academic Center, Emeq Hefer 40250, Israel}
\address{ }
\address{The Hadassah Institute, Jerusalem 91010, Israel}
\date{\today}

\begin{abstract}
\ \ \ The weak cosmic censorship conjecture asserts that spacetime
singularities that arise in gravitational collapse are always hidden
inside of black holes, invisible to distant observers. This
conjecture, put forward by Penrose more than four decades ago, is
widely believed to be one of the basic principles of nature.
However, a complete proof of this hypothesis is still lacking and
the validity of the conjecture has therefore remained one of the
most important open questions in general relativity. In this study
we analyze a gedanken experiment which is designed to challenge
cosmic censorship by trying to overcharge a Reissner-Nordstr\"om
black hole: a charged shell is lowered {\it adiabatically} into the
charged black hole. The mass-energy delivered to the black hole can
be red-shifted by letting the dropping point of the shell approach
the black-hole horizon. On the other hand, the electric charge of
the shell is not red-shifted by the gravitational field of the black
hole. It therefore seems, at first sight, that the charged shell is
not hindered from entering the black hole, overcharging it and
removing its horizon. However, in the present study we prove that
the exposure of a naked singularity to distant observers is actually
excluded due to the formation of a new (and {\it larger}) horizon
around the original black hole. Moreover, we shall prove that this
new horizon is already formed {\it before} the charged shell crosses
the original black-hole horizon. This result, which seems to have
been previously overlooked, guarantees the validity of the weak
cosmic censorship conjecture in this type of gedanken experiments.
\end{abstract}
\bigskip
\maketitle


The singularity theorems of Hawking and Penrose \cite{HawPen} reveal
that gravitational collapse from smooth initial conditions may
produce spacetime singularities, regions in which the known laws of
physics break down. The utility of general relativity in describing
gravitational phenomena in such extreme physical situations is
maintained by the cosmic censorship principle \cite{Pen,Haw1,Brady}.
The weak version of this hypothesis [the weak cosmic censorship
conjecture (WCCC)] asserts that spacetime singularities that arise
in gravitational collapse are always hidden inside of black holes
(behind event horizons), invisible to distant observers.

The cosmic censorship principle is essential for preserving the
predictability of Einstein's theory of gravity
\cite{Pen,Haw1,Brady}. In fact, the principle has become one of the
cornerstones of general relativity. However, a generic proof of the
conjecture is still lacking. Thus, the validity of this principle
has remained one of the most important open questions in general
relativity, see e.g.
\cite{Wald,Wald1,Sin,Clar,Vis,Price,His,KayWal,BekRos,Hub,QuiWal,Hod1,HodPir,Hod2,ForRom1,
ForRom2,MatSil,Hodlet,Hodplb,Elin,Tedso,Mar,Hodcc,Saa,Lee,Poiss} and
references therein.

According to the WCCC, the destruction of a black-hole event horizon
is ruled out because such process would expose the inner black-hole
singularity to distant observers. For this reason, any physical
process which is aimed to remove the black-hole horizon is expected
to fail. For the advocates of the cosmic censorship conjecture the
task remains to find out how such candidate processes eventually
fail to remove the black-hole horizon.

One of the earliest attempts to remove the horizon of a black hole
is due to Wald \cite{Wald} who tried to over-charge a maximally
charged Reissner-Nordstr\"om (RN) black hole by dropping into it a
charged test particle whose charge-to-mass ratio is larger than
unity. According to the uniqueness theorems
\cite{un1,un2,un3,un4,un5}, all spherically-symmetric black-hole
solutions of the Einstein-Maxwell equations are uniquely described
by the RN metric
\begin{equation}\label{Eq1}
ds^2=-\Big(1-{{2M}\over{r}}+{{Q^2}\over{r^2}}\Big)dt^2+\Big(1-{{2M}\over{r}}+{{Q^2}\over{r^2}}\Big)^{-1}dr^2
+r^2d\Omega^2\
\end{equation}
which is characterized by two conserved parameters: the
gravitational mass $M$ and the electric charge $Q$. The black-hole
(event and inner) horizons are located at
\begin{equation}\label{Eq2}
r_{\pm}=M\pm (M^2-Q^2)^{1/2}\  .
\end{equation}
Thus, a black-hole solution must satisfy the relation
\begin{equation}\label{Eq3}
Q^2\leq M^2\  .
\end{equation}
Maximally charged (extremal) black holes are the ones which saturate
the condition (\ref{Eq3}). The RN spacetime with $M^2<Q^2$ does not
contain an event horizon and is therefore associated with a naked
singularity rather than a black hole.

Wald \cite{Wald} considered the specific case of a charged particle
which starts falling towards the black hole from spatial {\it
infinity}. Thus, the particle's energy-at-infinity was larger than
(or equal to) its rest mass. It was shown \cite{Wald} that this
particular attempt to over-charge the black hole fails due to the
coulomb potential barrier which surrounds the charged black hole. A
similar gedanken experiment was studied by Hubeny \cite{Hub} who
tried to over-charge a near-extremal RN black hole using a charged
imploding shell. It was found \cite{Hub} that this attempt to remove
the black-hole horizon also fails --- the repulsive coulomb
interaction between the black hole and the shell and the coulomb
self-repulsion of the shell itself both prevent the shell from
over-charging the black hole.

In the present study we shall analyze a more dangerous version (from
the point of view of the WCCC) of the overcharging gedanken
experiment. This version consists a charged object which is lowered
{\it slowly} into the black hole. In this scenario, the energy
delivered to the black hole [the part contributed by the rest mass
of the object, see Eq. (\ref{Eq4}) below] can be {\it red-shifted}
by letting the dropping point of the object approach the black-hole
horizon. On the other hand, the electric charge of the object is not
red-shifted by the black-hole gravitational field. The
charge-to-energy ratio of a slowly-descending charged object is
therefor larger than the corresponding charge-to-energy ratios of
the free falling (from infinity) objects considered in the original
gedanken experiments \cite{Wald,Hub}. Thus, the present version of
the overcharging gedanken experiment poses a stronger challenge to
the cosmic censorship conjecture.

We consider a spherical charged shell of rest mass $m$ and electric
charge $q$ concentric with a charged RN black hole of mass $M$ and
electric charge $Q$. Our aim is to challenge the validity of the
WCCC in the most dangerous situation --- when the charge-to-energy
ratio of the shell is as large as possible. We shall therefore
consider a shell which is lowered slowly towards the charged black
hole. Our plan is to lower the shell adiabatically (that is, with an
infinitesimally small radial velocity) all the way down to the
black-hole horizon. The mass-energy of the shell would then be
red-shifted by the gravitational field of the black hole. This
adiabatic process would therefore minimize the energy which is
delivered to the black hole (for a given value of the shell's
electric charge).

The total energy of the shell in the black-hole spacetime is given
by \cite{Hub,NoteHub}
\begin{equation}\label{Eq4}
{\cal
E}(R)=m\Big(1-{{2M}\over{R}}+{{Q^2}\over{R^2}}\Big)^{1/2}+{{qQ}\over{R}}+{{q^2}\over{2R}}
-{{m^2}\over{2R}}\  ,
\end{equation}
where $R$ is the radius of the shell.
Each term on the r.h.s. of Eq. (\ref{Eq4}) has a clear physical
interpretation:
\begin{itemize}
\item{The first term represents the energy associated
with the shell's rest mass (red-shifted by the gravitational field
of the black hole).}
\item {The second term represents the electrostatic
interaction of the charged shell with the charged black hole.}
\item{The third term represents the electrostatic self-energy of the
charged shell.}
\item{The fourth term represents the gravitational self-energy of
the shell.}
\end{itemize}

The generalized Birkhoff's theorem implies that the spacetime inside
the shell is described by the RN metric (\ref{Eq1}) with parameters
$M$ and $Q$, whereas the spacetime outside the shell in described by
the RN metric (\ref{Eq1}) with total mass $M+{\cal E}(R)$ and total
electric charge $Q+q$.

Suppose the charged shell is indeed lowered adiabatically all the
way down to the original black-hole horizon. In this case, the
mass-energy of the shell is completely red-shifted by the
gravitational field of the black hole and the energy and electric
charge which are delivered to the black hole are given by: $\Delta
M={\cal E}(R=r_+)=qQ/r_++q^2/2r_+-m^2/2r_+$ and $\Delta Q=q$,
respectively. If this scenario would have been possible, then
charged shells with
\begin{equation}\label{Eq5}
r_+-Q-m<q<r_+-Q+m\
\end{equation}
could overcharge the black hole [that is, could violate the
black-hole condition (\ref{Eq3})], thereby violating the WCCC.

However, we shall now prove that the shell {\it cannot} be lowered
adiabatically all the way down to the original black-hole horizon.
In particular, we shall prove that a new (and {\it larger}) horizon
is formed outside the original black-hole horizon (that is, outside
$r_+$) already {\it before} the charged shell crosses the original
black-hole horizon. The characteristic condition for the formation
of a new horizon which engulfs {\it both} the original black hole
and the descending charged shell is
\begin{equation}\label{Eq6}
1-{{2[M+{\cal E}(R)]}\over{R}}+{{(Q+q)^2}\over{R^2}}=0\  .
\end{equation}
Substituting (\ref{Eq4}) into (\ref{Eq6}), one finds that a new
horizon is formed when the radius of the shall reaches the limiting
value
\begin{equation}\label{Eq7}
R\to r_{\text{NH}}\equiv M+(M^2-Q^2+m^2)^{1/2}\  .
\end{equation}

It is important to emphasize that the new horizon is formed {\it
outside} \cite{Noteout} the original black hole already {\it before}
the shell crosses the original horizon. The formation of the new
shielding horizon outside the original black hole prevents the
exposure of the inner singularity to distant observers. The newly
formed horizon therefore guarantees the validity of the WCCC in this
gedanken experiment.

We note that the radius (\ref{Eq7}) is the smallest possible radius
of such newly formed horizons: a charged shell with a non-vanishing
radial momentum has an energy which is larger than the one given by
(\ref{Eq4}) and would therefore form a larger horizon [that is, even
before \cite{Notebef} reaching the radius (\ref{Eq7}).]

It is worth reexamining a related gedanken experiment designed by
Bekenstein and Rosenzweig \cite{BekRos} to challenge cosmic
censorship \cite{Notedan}: suppose there exist {\it two} different
types of local charges (for example, electric and magnetic charges).
A RN spacetime with two different types of charges, $Q\in U(1)$ and
$K\in U'(1)$, can have an event horizon only if
\begin{equation}\label{Eq8}
Q^2+K^2\leq M^2\  .
\end{equation}

Suppose the original black hole possesses a $U(1)$ charge $Q$ but no
$U'(1)$ charge. Thus, the original black hole is not endowed with a
$U'(1)$ gauge field and an approaching shell of charge $k\in U'(1)$
encounters no electrostatic repulsion from the $U(1)$-charged black
hole [see Eq. (\ref{Eq9}) below]. Thus, the charge-to-energy ratio
of the shell is larger than the corresponding ratio considered in
the former gedanken experiment [with only one type of local $U(1)$
charge]. Hence, this type of gedanken experiment seems to pose a
greater challenge to the WCCC.

Bekenstein and Rosenzweig \cite{BekRos} considered a charged shell
which starts falling towards the black hole from spatial {\it
infinity}. They then concluded that the coulomb self-repulsion of
the shell is sufficient to guarantee the validity of the WCCC in
their version of the gedanken experiment \cite{BekRos}.

However, in our version of the gedanken experiment (which is more
challenging from the point of view of the WCCC) the shell is lowered
{\it adiabatically} towards the original black hole. As discussed
above, in this case the mass-energy of the shell is red-shifted by
the gravitational field of the black hole. As a consequence, the
charge-to-energy ratio of the shell is larger than the corresponding
charge-to-energy ratio considered in \cite{BekRos}. The total energy
of the shell in the black-hole spacetime is now given by
\begin{equation}\label{Eq9}
{\cal
E}(R)=m\Big(1-{{2M}\over{R}}+{{Q^2}\over{R^2}}\Big)^{1/2}+{{k^2}\over{2R}}
-{{m^2}\over{2R}}\  .
\end{equation}
Note, in particular, that the repulsion term $qQ/R$ that appeared in
(\ref{Eq4}) is absent now.

Suppose the $U'(1)$-charged shell is indeed lowered adiabatically
all the way down to the original horizon of the $U(1)$-charged black
hole. In this case, the mass-energy of the shell is totally
red-shifted by the gravitational field of the black hole and the
energy and electric charges which are delivered to the black hole
are given by: $\Delta M={\cal E}(R=r_+)=k^2/2r_+-m^2/2r_+, \Delta
Q=0$, and $\Delta K=k$. If this scenario would have been possible,
then charged shells with
\begin{equation}\label{Eq10}
m^2+2r_+(r_+-M-m)<k^2<m^2+2r_+(r_+-M+m)
\end{equation}
could overcharge the black hole [that is, could violate the
black-hole condition (\ref{Eq8})], thereby violating the WCCC.

However, it is easy to verify that a new and larger horizon is
formed {\it outside} the original black-hole horizon (that is,
outside $r_+$) already {\it before} the charged shell crosses the
original black-hole horizon. The characteristic condition for the
formation of a new horizon (which again engulfs both the original
black hole and the descending charged shell) is
\begin{equation}\label{Eq11}
1-{{2[M+{\cal E}(R)]}\over{R}}+{{Q^2+k^2}\over{R^2}}=0\  .
\end{equation}
Substituting (\ref{Eq9}) into (\ref{Eq11}), one finds that a new
horizon is formed when the radius of the shall reaches the limiting
value $r_{\text{NH}}$ given by Eq. (\ref{Eq7}). We therefore recover
our previous conclusion --- the new shielding horizon, which is
formed outside the original black hole, prevents the exposure of the
inner singularity to distant observers. Cosmic censorship is
therefore respected.

In summary, we have analyzed a gedanken experiment that was designed
to challenge the cosmic censorship conjecture by trying to
overcharge a black hole: a charged shell was lowered {\it
adiabatically} towards a charged Reissner-Nordstr\"om black hole.
The charge-to-energy ratio of the shell was made as large as
possible by red-shifting the energy associated with the rest mass of
the shell. Thus, the present gedanken experiment is more challenging
(from the point of view of the cosmic censorship conjecture) than
former gedanken experiments considered in \cite{Wald,BekRos,Hub}.

We have proved that when the shell approaches the original
black-hole horizon (but has {\it not} yet crossed it!), a new and
larger horizon is formed which engulfs {\it both} the original black
hole and the descending charged shell, see Eq. (\ref{Eq7}).

The formation of the new horizon outside the original black hole
{\it before} the shell crosses the original horizon, a fact which
seems to have been previously overlooked, prevents the exposure of
the inner singularity to distant observers. The newly formed
shielding horizon therefore guarantees the validity of the WCCC in
this type of gedanken experiments.

\bigskip
\noindent
{\bf ACKNOWLEDGMENTS}
\bigskip

This research is supported by the Carmel Science Foundation. I thank
Yael Oren, Arbel M. Ongo and Ayelet B. Lata for stimulating
discussions.


\begin{thebibliography}{99}

\bibitem{HawPen} S. W. Hawking and R. Penrose, Proc. R. Soc. London
A {\bf 314}, 529 (1970).

\bibitem{Pen} R. Penrose, Riv. Nuovo Cimento {bf 1}, 252 (1969); in
{\it General Relativity, an Einstein Centenary Survey}, edited by S.
W. Hawking and W. Israel (Cambridge University Press, Cambridge,
England, 1979).

\bibitem{Haw1} S. W. Hawking, Phys. Rev. D {\bf 14}, 2460 (1976).

\bibitem{Brady} P. R. Brady, I. G. Moss, and R. C. Myers, Phys. Rev.
Lett. {\bf 80}, 3432 (1998).

\bibitem{Wald} R. Wald, Ann. Phys. (N. Y.) {\bf 82}, 548 (1974).

\bibitem{Wald1} R. M. Wald, ``Gravitational Collapse and Cosmic
Censorship", e-print gr-qc/9710068.

\bibitem{Sin} T. P. Singh, ``Gravitational Collapse, Black Holes and
Naked Singularities", e-print gr-qc/9805066.

\bibitem{Clar} C. J. S. Clarke, Class. and Quant. Grav. {\bf 11}, 1375 (1994).

\bibitem{Vis} C. V. Vishveshwara, Phys. Rev. D {\bf 1}, 2870 (1970).

\bibitem{Price} R. Price, Phys. Rev. D {\bf 5}, 2419 (1972); {\bf
5}, 2439 (1972).

\bibitem{His} W. A. Hiscock, Ann. Phys. (N.Y.) {\bf 131}, 245
(1981).

\bibitem{KayWal} B. S. Kay and R. M. Wald, Classical and Quantum
Gravity {\bf 4}, 893 (1987).

\bibitem{BekRos} J. D. Bekenstein and C. Rosenzweig, Phys. Rev. D
{\bf 50}, 7239 (1994).

\bibitem{Hub} V. E. Hubeny, Phys. Rev. D {\bf 59}, 064013 (1999).

\bibitem{QuiWal} T. C. Quinn and R. M. Wald, Phys. Rev. D {\bf 60}, 064009 (1999).

\bibitem{Hod1} S. Hod, Phys. Rev. D {\bf 60}, 104031 (1999) [arXiv:gr-qc/9907001].

\bibitem{HodPir} S. Hod, e-print gr-qc/9908004; S. Hod and T. Piran,
Gen. Relativ. Gravit. {\bf 32}, 2333 (2000).

\bibitem{Hod2} S. Hod, Phys. Rev. D {\bf 66}, 024016 (2002)
[arXiv:gr-qc/0205005].

\bibitem{ForRom1} L. H. Ford and T. A. Roman, Phys. Rev. D {\bf 41}, 3662 (1990).

\bibitem{ForRom2} L. H. Ford and T. A. Roman, Phys. Rev. D {\bf 46},
1328 (1992).

\bibitem{MatSil} G. E. A. Matsas and A. R. R. da Silva, Phys. Rev.
Lett. {\bf 99}, 181301 (2007).

\bibitem{Hodlet} S. Hod, Phys. Rev. Lett. {\bf 100}, 121101 (2008)
[arXiv:0805.3873].

\bibitem{Hodplb} S. Hod, Phys. Lett. B {\bf 668}, 346 (2008)
[arXiv:0810.0079].

\bibitem{Elin} C. Eling and J. D. Bekenstein, Phys. Rev. D {\bf 79}, 024019
(2009).

\bibitem{Tedso} T. Jacobson, T. P. Sotiriou, Phys. Rev. Lett. {\bf 103}, 141101 (2009).

\bibitem{Mar} M. Bouhmadi-Lopez, V. Cardoso, A. Nerozzi, and J. V.
Rocha, Phys. Rev. D {\bf 81}, 084051 (2010).

\bibitem{Hodcc} S. Hod, Phys. Lett. B {\bf 693}, 339 (2010)
[arXiv:1009.3695].

\bibitem{Saa} A. Saa and R. Santarelli, Phys. Rev. D {\bf 84},
027501 (2011).

\bibitem{Lee} B. Gwak and B. H. Lee, Phys. Rev. D {\bf 84}, 084049
(2011).

\bibitem{Poiss} P. Zimmerman, I. Vega, E. Poisson, and R. Haas,
arXiv:1211.3889.

\bibitem{un1} W. Israel, Phys. Rev. {\bf 164}, 1776 (1967); Commun.
Math. Phys. {\bf 8}, 245 (1968).

\bibitem{un2} B. Carter, Phys. Rev. Lett. {\bf 26}, 331 (1971).

\bibitem{un3} S. W. Hawking, Commun. Math. Phys. {\bf 25}, 152 (1972).

\bibitem{un4} D. C. Robinson, Phys. Rev. D {\bf 10}, 458 (1974); Phys. Rev.
Lett. {\bf 34}, 905 (1975).

\bibitem{un5} J. Isper, Phys. Rev. Lett. {\bf 27}, 529 (1971).

\bibitem{NoteHub} Here we have solved the equation of motion of the
shell [see Eq. (35) of \cite{Hub}]: $\sqrt{g_{\text{in}}(r)+\dot
R^2}-\sqrt{g_{\text{out}}(r)+\dot R^2}=-m/r$ with $\dot R=0$. Here
$g_{\text{in}}(r)=1-2M/r+Q^2/r^2$ and $g_{\text{out}}(r)=1-2(M+{\cal
E})/r+(Q+q)^2/r^2$.

\bibitem{Noteout} That is, $r_{\text{NH}}>r_+$ [see
Eqs. (\ref{Eq2}) and (\ref{Eq7})].

\bibitem{Notebef} For example, a shell
which starts falling from rest towards the black hole from spatial
infinity would form a new horizon when reaching
$r_{\text{NH}}=M+m+[(M+m)^2-(Q+q)^2]^{1/2}$.

\bibitem{Notedan} Our version of the gedanken experiment would be
more challenging (from the point of view of the WCCC) than the one
considered in \cite{BekRos}.

\end{thebibliography}
\end{document}